\documentstyle[aps,twocolumn,graphicx,floats,prl,xspace,amsmath]{revtex}

\DeclareMathOperator{\sn}{sn}
\DeclareMathOperator{\cn}{cn}
\DeclareMathOperator{\dn}{dn}
\DeclareMathOperator{\cd}{cd}

\newcommand{\ncharged}{\Gamma_{\mathrm{SiO^-}}}
\newcommand{\nneutral}{\Gamma_{\mathrm{SiOH}}}
\newcommand{\ntotal}{\Gamma}
\newcommand{\pactivity}{\left[\mathrm{H}^+\right]_0}
\newcommand{\pH}{\mathrm{pH}}
\newcommand{\pK}{\mathrm{pK}}
\newcommand{\micronsq}{\mu {\rm m}^2 \xspace}
\newcommand{\micron}{$\mu$m\xspace}
\newcommand{\dd}{\hspace{2pt}d} 
\newcommand{\eps}{\varepsilon}

\newcommand{\s}{\sigma}

\begin{document}

\wideabs{
\title{The Charge of Glass and Silica Surfaces}

\author{Sven H. Behrens and David G. Grier}

\address{Dept. of Physics, James Franck Institute, and
Institute for Biophysical Dynamics\\ 
The University of Chicago, Chicago, IL 60637}

\date{\today}

\maketitle

\begin{abstract}
We present a method of calculating the electric charge density of glass and 
silica surfaces in contact with aqueous electrolytes 
for two cases of practical relevance 
that are not amenable to standard techniques: surfaces of low specific area
at low ionic strength and surfaces interacting strongly with a second anionic 
surface.
\end{abstract}
} 

\section{Introduction}
Ionization processes in aqueous solutions have long been a point of central
interest to physical chemistry. 
Much progress has been made in recent years in understanding 
the charging properties of such different entities as small molecules, 
polyelectrolytes and various kinds of interfaces \cite{borkovec01}.

In this context, silica and silicate glass surfaces immersed in 
water are known to
acquire a negative surface charge density, primarily
through the dissociation of terminal silanol groups.
The degree of dissociation and thus the surface charge density
results from an equilibrium between counterions at the glass surface
and free ions in the bulk electrolyte.
Experimentally, this type of equilibrium and its dependence on the solution
conditions can be studied by potentiometric acid-base titrations 
on colloidal dispersions of non-porous silica particles \cite{iler79}.
This technique actually measures the volume concentration of 
protons transferred between the surfaces and the solution.  
In order for the surfaces to accommodate a sufficient amount of charge,  
the electrostatic interaction between the surface sites must be 
screened at least partially by added salt ions, and/or the available surface area must be large.

These constraints can be relaxed to some degree by
resorting to alternative techniques like microelectrophoresis
\cite{kosmulski92,sanders95}, streaming potential measurements \cite{gu00},
conductometry \cite{yamanaka97}
and electroacoustic methods \cite{rosen91,rasmusson97}.
All of these methods, however, rely heavily on approximate models for electrostatic or hydrodynamic processes in the interfacial region, introducing
uncertainties that are difficult to estimate. 
We are not aware of any way to measure directly the surface charge of silica
in a solution of very low ionic strength.

Theoretical studies of low ionic strength solutions typically deal with
dense colloidal and macroionic systems and consider the regime of ``no salt'',
``low salt'', or ``counterions only''. Here the defining assumption is that the 
overall ionic strength is due predominantly to the particles or macroions and 
the compensating counterions in solution, whereas the concentration of any 
additional ions is negligible. For colloidal dispersions this assumption 
is again legitimate if the specific surface area carrying the colloidal charge
is large.

A series of recent experiments has spurred interest in the charge on glass and silica surfaces of low specific area in pure water, {\em i.e.} systems for which the usual picture of the ``no salt'' regime does not apply.  
For example, interaction measurements using digital video microscopy 
and optical trapping suggest that highly charged latex spheres
may experience an anomalous long-ranged attraction
when confined by charged glass walls
\cite{kepler94,carbajaltinoco96,crocker96,larsen97},
contrary to the predictions of Poisson-Boltzmann theory 
\cite{sader99,neu99,sader00}.
In one particular case \cite{larsen97}, the attraction
appears to result from a hydrodynamic interaction driven by
the spheres' electrostatic repulsion from a nearby wall
\cite{squires00}.
This explanation hinges on the heretofore untested
assumption that the glass
wall carries an effective charge density of
$-2000 \pm 200~e/\micronsq$, where $e$ is the elementary charge.
Such hydrodynamic coupling cannot explain 
the like-charge attractions measured for
spheres confined between two charged glass walls
\cite{kepler94,carbajaltinoco96,crocker96}.
How the walls' charge influence colloidal electrostatic
interactions is not yet resolved, in part because of open
questions regarding the charging state of the glass.
We recently reported that the pair interaction of silica spheres
remains monotonically repulsive even in the presence of a charged glass
wall \cite{behrens01}.
The spheres' effective surface charge density of $- 700 \pm 150~e/\micronsq$
extracted from these measurements is considerably
smaller than the value posited in Ref.~\cite{squires00} for a compact
glass surface.

Because a silica surface's charge density depends on the local chemical
environment, it necessarily varies with proximity to other
charge-carrying surfaces. The interpretation of typical
particle deposition experiments \cite{elimelech95} and force measurements by 
atomic force microscopy (AFM) \cite{butt91,ducker92} or total internal 
reflection microscopy (TIRM) \cite{prieve99} for example, 
is complicated by the fact that the charge densities of the substrate 
and the probe are a function of their separation, a phenomenon known as
``charge regulation''.
Since local properties of the enclosed solution rather than bulk 
properties determine the charging state, a naive use of 
charging data from bulk measurements can lead to errors.

In the following section
we discuss how the experimentally supported 1-pK Basic Stern Model for
silica surfaces may be used to calculate the elusive charge of glass plates 
and strongly diluted silica particles in deionized water. 
In the remainder of this paper we take advantage of a recently proposed 
theoretical treatment of charge regulation \cite{behrens99b,behrens99} to 
discuss the
the charge of a silica-like surface in close proximity to a second
anionic surface, 
which will be chosen, in view of the most common applications,
as either of the same type or of constant charge (like sulfate latex) or of
carboxylic nature (like carboxyl latex and many biological surfaces).

\section{Effective Charge of Glass and Silica in Deionized Solutions}
The principal mechanism by which glass and silica surfaces 
acquire a charge in contact
with water is the dissociation of silanol groups \cite{iler79}
\begin{equation}
  {\rm SiOH \rightleftharpoons SiO^- + H^+ . } 
  \label{dissociation}
\end{equation}
Further protonation of the uncharged group is expected only under extremely
acidic conditions \cite{hiemstra89,hiemstra96}
and will be disregarded. Similarly, we will not take into account the protonation of doubly coordinated ${\rm Si}_2-{\rm O}$ groups as these are 
generally considered inert \cite{hiemstra89}.

In addition to the hydronium or other counterions 
dissociated from the surface, the bulk electrolyte also includes
ions due to the autodissociation of
water; the latter can dominate the concentration of mobile ions 
if the ratio of surface area to solution volume is exceedingly small. 
Under these conditions, the charging state of the surface is controlled by the
ionic strength and pH of the bulk electrolyte,
just as in the general case of high salt concentrations.
We propose to take advantage of this similarity by
applying insights gained from
studies at high electrolyte concentrations 
to calculate the charge density at silica-water interfaces
with low surface area and 
with no added salt.

Whether the specific surface area is indeed small enough to warrant a ``high salt'' treatment, depends largely on the 
geometry of the considered experimental setup and has to be checked on a case-by-case basis. For some of the aforementioned interaction measurements \cite{larsen97,behrens01} this 
approach is appropriate, in other cases it may provide a very rough upper limit for the surface charge. If, on the other hand, 
the counterions due to the charged surfaces give a non-negligible contribution 
to the overall ionic strength, they have to be considered explicitly, for
instance within a cell model \cite{alexander84,gisler94}. 
Here, we concentrate on the former case of ``high salt'' and adopt the Basic Stern model \cite{westall80}, which has been shown to accurately describe titration data \cite{bolt57} obtained in the this regime
for nominally nonporous, fully hydrated 
silica particles \cite{hiemstra89,hiemstra96}.

Within the Basic Stern model the charge of silica
is regarded as localized entirely on the surface and arising
from a concentration $\ncharged$ of dissociated head groups \cite{behrens99b},
giving rise to the surface charge density 
\begin{equation}
  \sigma  =  - e \ncharged. 
  \label{chargedensity}
\end{equation}
Under normal conditions, only a fraction of the total concentration,
\begin{equation}
  \ntotal  =  \ncharged + \nneutral,
  \label{def_gamma}
\end{equation}
of chargeable sites dissociate.
The relevant mass action law for the deprotonation
reaction, Eq.~(\ref{dissociation}),
\begin{equation}
  \frac{\pactivity  \ncharged}{\nneutral}
  = 10^{-\pK}\, \mathrm{Mol/l},
  \label{massactionlaw}
\end{equation}
is characterized by the logarithmic dissociation constant, pK, and
accounts for the influence of
the surface's electrostatic potential, $\psi_0$,
through the surface activity of protons,
\begin{equation}
\pactivity  = 
\left[\mathrm{H}^+\right]_{\mathrm b} \exp\left(-\beta e\psi_0\right).
\label{hsurf}
\end{equation}
Here, $\left[\mathrm{H}^+\right]_{\mathrm b} = 10^{-\pH}\, \mathrm{Mol/l}$ is the bulk activity of protons,
and $\beta^{-1}=k_BT$ denotes the thermal energy.
The dissociation constant is an inherent property of the silicate-water
interface and is estimated to be $\pK = 7.5$ on the basis
of a surface complexation model \cite{hiemstra89}.

As counterions dissociate from the surface, they form a diffuse cloud 
of charge within the electrolyte.
The Basic Stern model treats the counterions as being
separated from the surface by a thin Stern layer
across which the electrostatic potential drops linearly from 
is surface value, $\psi_0$, to a value $\psi_d$ called the diffuse layer potential \cite{westall80,behrens99b}.
This potential drop is characterized by the Stern layer's phenomenological 
capacity,
\begin{equation}
C = \frac{\sigma}{\psi_0 - \psi_d} 
\label{capacity}
\end{equation}
This capacity, $C$, reflects the structure of the silicate-water interface
and should vary little with changes in surface geometry or
electrolyte concentration.
Titration data on colloidal silica \cite{bolt57} are consistent
with $C = 2.9~\mathrm{F/m^2}$ \cite{hiemstra89}.

Eqs.~(\ref{chargedensity}--\ref{capacity}) can be solved 
for the diffuse layer potential
as a function of the charge density on the interface:
\begin{equation}
  \psi_d (\sigma) =  \frac{1}{\beta e} \, \ln\frac{-\sigma}{e\Gamma + \sigma}
  - (\pH - \pK) \frac{\ln 10}{\beta e} - \frac{\s}{C} .
  \label{eq:psichem}
\end{equation}
This relation
reflects the chemical
nature of the interface and its charging process.

Another functional dependence follows from
the distribution of mobile charges in the solution. 
If the latter is 
described by the Poisson-Boltzmann equation (PB),
then the charge of an isolated, flat surface satisfies the Grahame equation
\begin{equation}
  \sigma(\psi_d) = \frac{\eps\kappa}{2\pi\beta e} \,
  \sinh\left(\frac{\beta e \psi_d}{2}\right).
  \label{eq:grahame}
\end{equation}
Here, $\eps$ is the permittivity of the solution and $\kappa^{-1}$
the Debye screening length given by 
$\kappa^2 = 4\pi\beta e^2 n /\eps$, where $n$ is
the total concentration of small ions, all of which are
assumed to be monovalent.
The generalization of Eq.~(\ref{eq:grahame})
to account for a curvature of radius $a$,
\begin{equation}
  \sigma(\psi_d) =
  \frac{\eps\kappa}{2\pi\beta e}
  \left[\sinh\left(\frac{\beta e\psi_d}{2}\right) + 
    \frac{2}{\kappa a} \, \tanh\left(\frac{\beta e\psi_d}{4}\right)
  \right],
  \label{eq:low}
\end{equation}
is known to give the surface charge density to within 5\% for
$\kappa a \ge 0.5$ and any surface potential \cite{russel89}.

Combining Eq.~(\ref{eq:psichem}) with Eq.~(\ref{eq:grahame}) or (\ref{eq:low})
yields self-consistent values for the surface charge density, $\sigma$,
and the diffuse layer potential, $\psi_d$.
These values characterize the equilibrium of bound and mobile charges in 
the interfacial region,
but are not necessarily accessible experimentally, given the 
requirement of large surface areas for potentiometric titrations and the 
interpretive ambiguities inherent to other techniques.

Most measurements of interfacial interactions probe the electrostatic potential
$\psi$ at distances for which  $e\psi \le k_BT$. 
Under these conditions $\psi$ is described accurately by the linearized 
Poisson-Boltzmann equation, whose solution for a single flat surface has
the form $\psi(x)=\psi_{\rm eff} \exp(-\kappa x)$, where $x$ is the distance 
from the surface.
The effective surface potential $\psi_{\rm eff}$ in this experimentally
accessible regime is related
to the actual diffuse layer potential through
\cite{russel89}
\begin{equation}
\beta e \psi_{\rm eff}  =  
  4 \tanh\left(\frac{\beta e \psi_d }{4}\right).
\end{equation} 

Again, there is an approximate generalization for curved surfaces 
\cite{oshima82}:
\begin{equation}
  \beta e \psi_{\rm eff}  =  
  \frac{8 \tanh\left(\frac{\beta e\psi_d}{4}\right)}{
    1 + \left[
      1-\frac{1+2\kappa a}{(1+\kappa a)^2}
      \tanh^2\left(\frac{\beta e \psi_d}{4}\right)
    \right]^{1/2}}.
  \label{eq:peff}
\end{equation}
The associated effective charge density
can be obtained from
\begin{equation}
  \sigma_{\rm eff} =  \frac{\eps\kappa}{4\pi}\psi_{\rm eff}
  \left[1+\frac{1}{\kappa a}
  \right],
  \label{eq:seff}
\end{equation}
which is just the linearization of Eq.~(\ref{eq:low}).
This effective charge density characterizes essentially all
of the recent measurements of
electrostatic interactions between well-separated charged surfaces.

The effective charge's relevance to experimental observations
is based in the popularity of the linear superposition approximation
for estimating the interaction energy, $u(h)$, between
two charged spheres of radii $a_1$ and $a_2$ as a function
of their surface-to-surface separation $h$.
In this approximation \cite{bell70},
\begin{equation}
  u(h)  =  \frac{1}{\eps}
  \left( \frac{\sigma_1 a_1^2}{1+\kappa a_1} \right)
  \left( \frac{\sigma_2 a_2^2}{1+\kappa a_2} \right)
  \frac{\exp(-\kappa h)}{a_1+a_2+h}, 
  \label{eq:screenedC}
\end{equation}
where $\sigma_1$ and $\sigma_2$ should be understood 
to be effective surface charge densities obtained from 
Eq.~(\ref{eq:peff}) 
rather than the bare charge densities from 
Eqs.~(\ref{eq:psichem}) and (\ref{eq:low}).
Using the effective surface charge densities implicitly accounts for
overexponential decay of the electrostatic 
potential near the surfaces that follows from the nonlinearity
of the Poisson-Boltzmann equation.
The interaction between a sphere and a
planar wall is obtained by taking the limit of one infinite radius in 
Eq.~(\ref{eq:screenedC}).

\begin{figure}[t!]
  \centering
  \includegraphics[width=3in]{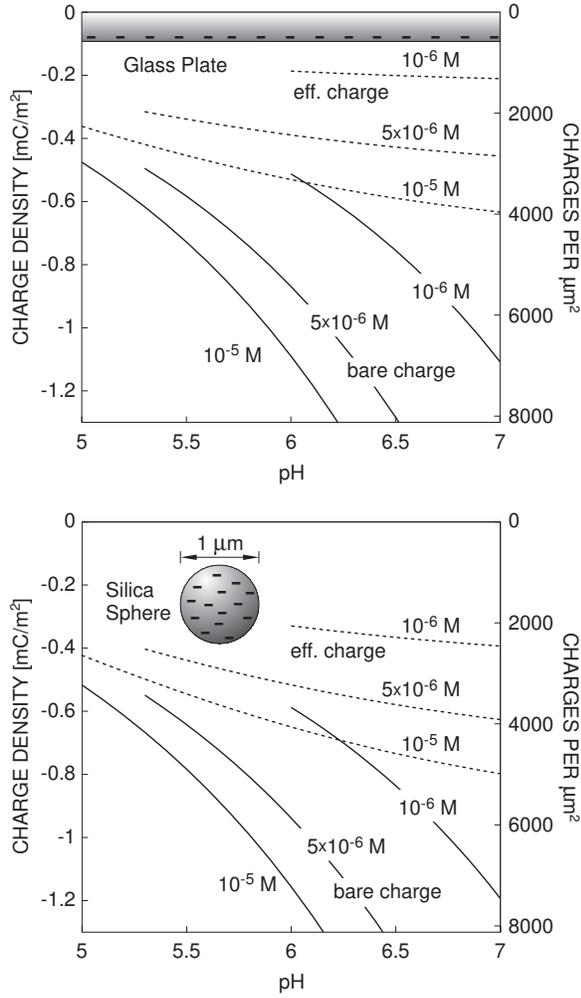}
  \vspace{1ex}
  \caption{The bare (full lines) and effective (dashed curves) charge
densities of a planar glass wall and a 1 micron silica sphere,
assuming a density $\Gamma=8$ ${\rm nm}^{-2}$ of chargeable sites, a
pK value of 7.5 for the silanol dissociation, and a Stern capacity of
2.9 ${\rm F}/{\rm m}^2$.}
  \label{fig:glass}
\end{figure}

Fig.~\ref{fig:glass} shows computed values for the bare and
effective charges of a planar silica surface and a 1~\micron-diameter
silica sphere for pH values between 7 and the lowest pH compatible with
an ionic strength of 1, 5, and 10 $\mu$Mol/l. 
These are reasonable values 
for deionized water under usual experimental conditions.
In addition to using $C = 2.9~\mathrm{F/m^2}$ and $\pK = 7.5$,
we have further assumed a total site density of 
$\Gamma = 8~\mathrm{nm}^{-2}$,
a commonly cited literature value for nonporous, fully hydrated
silica \cite{iler79}.
Although $\Gamma$ 
could vary widely depending on surface preparation,
the degree of protonation is determined mostly by the electrostatic 
interactions among the small fraction of charged surface sites, rather 
than the large number of neutral sites that $\Gamma$ accounts for, 
and so our results are quite insensitive to this parameter.
This robustness validates
our assumption that details in the structure of nonporous surfaces 
do not matter in the present context.
Indeed, the top graph of Fig.~\ref{fig:glass} 
also should represent the charging properties of a polished glass surface.
Note however that our arguments do not apply to some types of silica that are
believed to be
very porous and contain a much higher charge \cite{iler79}, 
the largest part of which seems located in the porous volume \cite{dekeizer98}
rather than on the surface.

Fig.~\ref{fig:glass} 
demonstrates that the effective charge densities in deionized
solutions do not depend as sensitively on pH as their ``bare'' counterparts, 
but that a significant variation with ionic strength persists.
The top part of the figure indicates that the value of 2000 
effective charges per
$\micronsq$ ($=-0.32~\mathrm{mC/m^2}$)
assumed by Squires and Brenner \cite{squires00} 
for a glass plate in contact with
deionized solution of $\kappa^{-1}=0.275$ \micron (i.e. an ionic strength of
$1.2 \times 10^{-6}$ M) is very reasonable. 
The confirmation of this previously very uncertain value 
provides vital support for their recent 
electro-hydrodynamic explanation of apparent attractions between
like-charged particles near a single glass wall.

In a study of the equilibrium 
interaction between few 1.58~\micron silica sphere at the bottom of a large
glass container filled with deionized water, we found an ionic strength
between $8.5 \times 10^{-7}$  and $1.1 \times 10^{-6}$ M
and a charge density between 550 and 830 $e/\micronsq$ from a 
fit of measured interaction energies to Eq.~(\ref{eq:screenedC}) 
\cite{behrens01}.
Comparison with 
Fig.~\ref{fig:glass} shows that these charge densities are a 
little below our expectation for isolated spheres, but have the 
right order of magnitude. 
The remaining difference can be explained by
the spheres' close proximity to the bottom wall of the 
glass container, as we describe in the next section.

\section{Charge Regulation of Anionic Surfaces}
Many experimental techniques to measure colloidal forces (AFM, TIRM, and the
surface force apparatus, for instance)
provide good resolution at very short distances.
Often the surface separations $h$ of interest are comparable to or 
smaller than the 
screening length and much smaller than the radii of curvature, $a_1$ and
$a_2$, of either surface.
In this situation, the surfaces may be regarded as locally flat, and the
Derjaguin approximation
\begin{equation}
  f(h) = 2\pi \frac{a_1 a_2}{a_1+a_2} W(h).
  \label{derjaguin}
\end{equation}
accurately expresses the
magnitude $f(h)$ of the acting force in terms of the interaction energy 
per unit area $W(h)$ for two parallel (thick) plates of the same separation.
This reduces the interaction problem to finding the energy $W(h)$
or the dividing pressure $\Pi(h)=-dW/dh $ resulting from a one-dimensional
distribution of mediating small ions.

On the other hand, the superposition of the noninteracting surfaces' 
electrostatic potential is no
longer warranted; nor can reliable results be expected from a solution of
the \emph{linearized} Poisson-Boltzmann equation, 
as it only captures the case of weak potentials (atypical for strongly
interacting surfaces) or, with renormalized charges, the asymptotic 
behavior for large separations. 

Instead, we consider the nonlinear PB equation, 
which in one dimension and for an excess of monovalent electrolyte ions reads
\begin{equation}
  \frac{d^2 \Psi}{d x^2}(x) =  \kappa^2 \, \sinh \Psi(x),
  \label{pb}
\end{equation}
where $\Psi = \beta e \psi$ 
is the dimensionless electrostatic potential
and $x$ the coordinate normal to the surfaces.
Applying this mean field formalism to more general electrolytes
shall not be discussed here, since neglected ion correlations and 
ion-specific interactions with the surfaces tend to complicate the case 
of polyvalent ions.

From Eq.~(\ref{pb}) it is clear that $\Psi(x)$ is a convex function for 
$\Psi \le 0$. The charge density on either surface (1 or 2) is given,
according to Gau{\ss}' law, by 
\begin{equation}
  \sigma_{1,2} =  -\left. \frac{\eps}{4\pi\beta e z}
    \frac{d\Psi}{dx}\right|_{1,2},
  \label{gauss}
\end{equation}
where the derivative is taken at the surface with respect to its outward
normal. It follows that between two negatively charged surfaces,
$\Psi(x)$ has a maximum $\Psi_{\rm m}$, which for identical surfaces lies 
exactly at the
midplane. 
Choosing generally the position of this maximum as the origin of our 
coordinate system (i.e. $\Psi(0)=\Psi_{\rm m}$), 
we can express the solution of Eq.~(\ref{pb}) as \cite{ninham71,behrens99}
\begin{equation}
  \Psi(x)  =  \Psi_{\mathrm m} + 2\ln \cd (u|m),
  \label{loesung}
\end{equation}
with 
\begin{equation}
  u = \frac{\kappa x}{2} \exp\left( -\Psi_{\mathrm m}/2 \right) 
\end{equation}
and
\begin{equation}
  m =  \exp\left( 2\Psi_{\mathrm m}\right),
\end{equation}
where $\cd(u|m)$ is a Jacobian elliptic function of argument $u$ and parameter
$m$ \cite{abramowitz72}.
The derivative is
\begin{equation}
  \frac{\dd\Psi}{\dd x}  =  
  \left( m^{3/4}-m^{-1/4}\right) \, \kappa \:
  \frac{\sn(u|m)}{\cn(u|m) \, \dn(u|m)},
  \label{deriv}
\end{equation}
where $\sn(u|m)$, $\cn(u|m)$, 
and $\dn=\cn/\cd$ are again Jacobian elliptic functions of 
the argument $u$ and parameter $m$ given above. Efficient numeric
implementations of these functions are readily available from mathematical 
libraries \cite{nag_math}. 

\subsection*{Equal Surfaces}
We shall measure the separation $h$ 
between the surfaces by the distance between the
head ends of the diffuse layer, i.e.,  $\Psi(h/2)=\beta e \psi_d$.
Evaluating Eqs.~(\ref{gauss}-\ref{deriv}) 
at $x=\pm h/2$
and combining them with the chemical boundary condition, 
Eq.~(\ref{eq:psichem}), provides an expression for the actual
surface charge density $\sigma$.
The second boundary condition, $d\Psi/dx=0$ at $x=0$, is already implied
in the solution for $\Psi(x)$, Eq.~(\ref{loesung}).
Note that for $\sigma(\psi_d,h)$ defined by Eqs.~(\ref{gauss}) through 
(\ref{deriv}), the long distance limit
$\lim_{h\to\infty} \sigma(\psi_d,h)$ 
is given by Eq.~(\ref{eq:grahame}), which we have
previously used for isolated surfaces and weakly interacting surfaces in the
superposition approximation. In general, $\sigma$ is a function of
$\psi_d$
as well as of the surface separation and cannot be expressed analytically.
As before, the electrostatic definition of $\sigma$, Eqs.~(\ref{gauss}) and
(\ref{deriv}),
depends parametrically on the Debye length, while
the chemical definition, Eq.~(\ref{eq:psichem}), 
depends only on pH and the surface chemical parameters,
$\Gamma$, pK, and $C$.

Technically, the combination of Eqs.~(\ref{eq:psichem}) and 
(\ref{gauss}-\ref{deriv})
results in a single transcendental equation for the midplane
potential $\Psi_{\rm m}$, whose numerical solution 
provides a very convenient alternative to numerical integration of the
differential equation (\ref{pb}) with nonlinear boundary conditions.

Solving for $\Psi_{\rm m}$ has the further advantage of immediately yielding 
the electrostatic force per unit area  
\begin{equation}
  \Pi  =   nk_BT \left( \cosh \Psi_{\rm m} -1 \right)
  \label{pressureeq},
\end{equation}
\emph{i.e.}~the excess osmotic pressure of small ions at the midplane where
the electric field and the associated Maxwell stress are zero.

\subsection*{Dissimilar Surfaces}
\begin{figure}[bthp]
  \centering
  \includegraphics[width=3in]{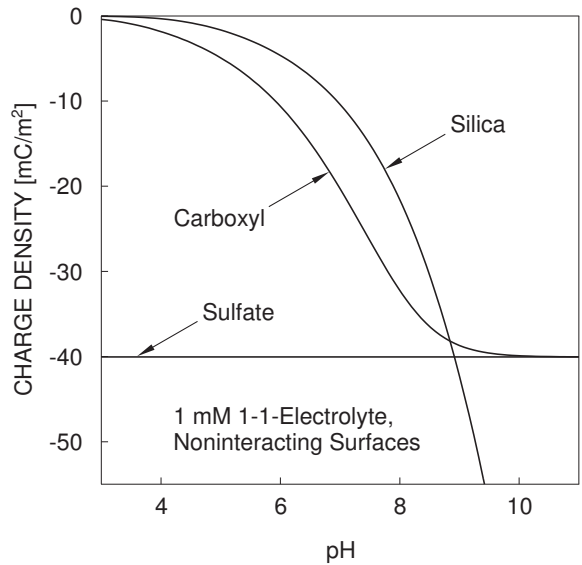}
  \vspace{1ex}
  \caption{
The charge density of surfaces with silanol, carboxyl, or sulfate head groups
as a function of pH.
Parameters
for the silica-like surface are as in Fig.~\ref{fig:glass}. For 
both the sulfate-
and the carboxyl bearing surface we have assumed a
density $\Gamma=0.25$ ${\rm nm}^{-2}$ 
of sites, all of which are constantly charged in the sulfate case;
further parameters of the carboxyl surface are 
a large (infinite) Stern capacity
and 
a dissociation pK of 4.9.}
  \label{fig:sigma_ph}
\end{figure}
\begin{figure}[b!]
  \centering
  \includegraphics[width=3in]{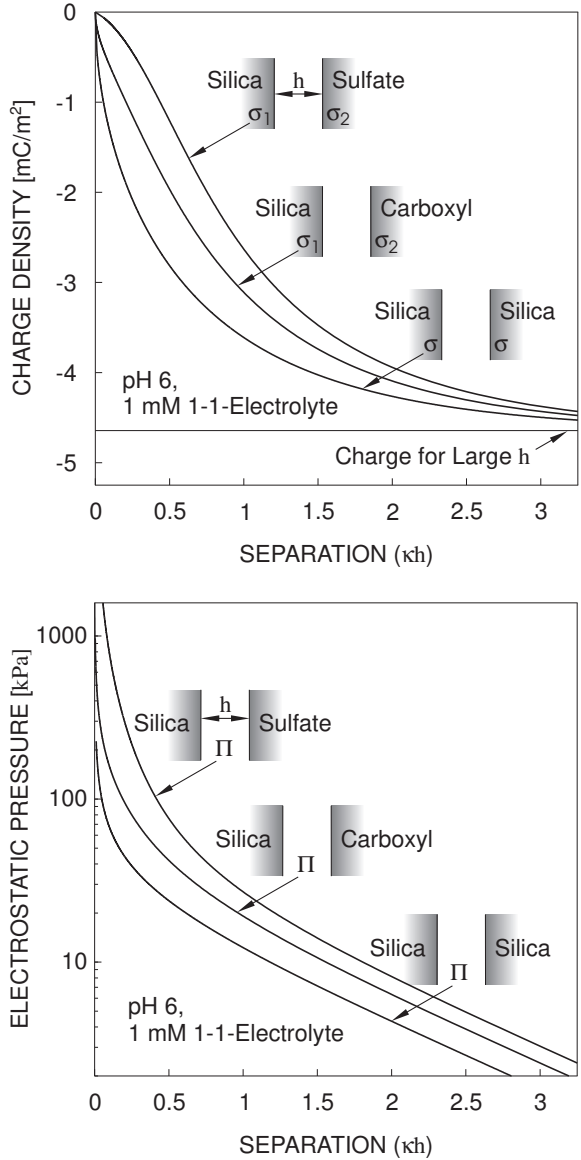}
  \vspace{1ex}
  \caption{The charge density of the silica surface and the force per unit 
area it experiences when interacting with any of the three 
presented types of surfaces at
pH 6 and an ionic strength of 1 mM ($\kappa^{-1}=9.6$ nm). Surfaces parameters
as in the previous figures.}
  \label{fig:interact}
\end{figure}
The procedure described before may be applied to 
negatively charged surfaces other than glass or silica as long as the 
chemically 
imposed charge-potential relation $\sigma(\psi_d)$ is modified to account 
for the surface properties of the considered material. 
Carboxylated latex for 
instance can be described in the same framework as silica, with a pK value
of 4.9 for the dissociation of the carboxyl surface groups 
(${\rm COOH \rightleftharpoons COO^- + H^+}$)
and a large Stern capacity $C$ (any value $C \gg 10$ amounting to a negligible
potential drop $|\psi_0-\psi_d|$ across the Stern layer) \cite{behrens00a}.
Sulfate latex, on the other hand, may be considered as having a constant
charge density ($\sigma(\psi_d)=-e\Gamma={\rm const.}$), because the 
strongly acidic sulfate groups are fully dissociated in all relevant
solution conditions.
Fig.~\ref{fig:sigma_ph} shows the predicted (and experimentally confirmed
\cite{hiemstra89,behrens00a,behrens98b})
charging behavior of the aforementioned
materials.
The site density of $\Gamma=0.25~\mathrm{nm^{-2}}$ chosen
for both the sulfate and the carboxyl surface lies in the typical range for
commercially available latex spheres and has also been cited as the 
density of carboxyl groups on the membrane of blood cells \cite{ninham71}.

A way to evaluate the interaction between two dissimilar surfaces starts by 
applying the described method for equal surfaces separately to both materials.
For each of these symmetric systems, 
one obtains the midplane potential $\Psi_{\rm m}$ 
and thus via Eq.~(\ref{loesung}) the full potential function $\Psi(x)$ 
associated with any given separation between equal plates.
Since $\Psi(x)$ is already fully determined by the value of 
$\Psi_{\rm m}=\Psi(0)$ and the requirement $d\Psi/dx|_{x=0}=0$, 
solutions $\Psi(x)$ of the Poisson-Boltzmann equation for different systems
are identical 
if they correspond to the same $\Psi_{\rm m}$ (\emph{i.e.}~the same pressure), 
the only difference being the surface separation $h$ for 
which they occur in the two systems.
A solution $\Psi(x)$ associated with a separation $h_1$ in one system and
with separation $h_2$ in the second system clearly serves as a solution
in a mixed system with a surface of type 1 at $x=-h_1/2$ and a surface
of type 2 at $x=+h_2/2$. Moreover, the separation $h=(h_1+h_2)/2$ at which 
this solution occurs in the mixed system is unique, because
the pressure is a monotonic function of separation in the symmetric systems
and can thus be inverted to give the two separation functions 
$h_1(\Psi_{\rm m} )$
and $h_2(\Psi_{\rm m})$.
Our strategy 
therefore consists of computing $\Psi_{\rm m}$ for all
separations of interest in the symmetric systems 1 and 2, finding the
separations 
$h_1(\Psi_{\rm m})$, $h_2(\Psi_{\rm m} )$ by inversion, and finally
inverting their arithmetic mean 
$h(\Psi_{\rm m})=[h_1(\Psi_{\rm m})+h_2(\Psi_{\rm m}) ]/2$ to obtain 
$\Psi_{\rm m}(h)$ and all the 
ensuing properties of interest in the mixed system.

Some results of this type are shown in Fig.~\ref{fig:interact}, where we have
plotted the charge density of a glass or silica surface and 
the electrostatic pressure as it interacts
with either its own kind or with a surface of the carboxyl or the sulfate
type.
At the chosen 
ionic strength of 1 mM and pH 6,
the charge of the silica surface is seen to 
deviate significantly from its value in isolation (horizontal line and 
Fig.~\ref{fig:sigma_ph}) up to separations of several screening lengths. 
Moreover, the nature of the second surface also has a profound effect not only 
on the strength of the interaction, but also on the charging state of the 
silica. While all anionic surfaces will reduce the effective charge on 
silica upon approach, the rate at which they do so strongly depends on the
amount and variability of their own charge. 
Neither of these dependencies are usually considered in 
the discussion of interaction measurements. 
A recent attempt to determine these
ionization properties of silica experimentally with atomic
force microscopy \cite{zhmud98} has been limited to symmetric surfaces, and
relies on model assumptions both for the charge regulation and 
for the strong van der Waals forces at short surface separations.

\section{Conclusions}
We have seen that the apparent charge on glass and 
silica surfaces of low specific 
area in pure water can be understood in terms of a simple model that was
originally developed and tested for high electrolyte concentrations.
Model predictions for effective charge densities compare favorably with
interaction experiments on highly diluted silica spheres in deionized 
water \cite{behrens01}. They also support a new kinematic explanation of 
spurious
long range attractions between like-charged particles near a single glass
wall \cite{squires00}.

The regulated charge of silica and glass surfaces near contact with
a second anionic surface, as well as the strength of the interaction, has been
calculated from an exact solution of the nonlinear Poisson-Boltzmann 
equation. In commonly encountered solution conditions,
the charge regulation of silica
was found to be effective at separations well beyond a Debye length. 
It also proves very sensitive to the chemical nature of the 
opposing surface. Although the additional presence of 
van der Waals forces makes a quantitative measurement of these
effects difficult, they should certainly be accounted for in the 
interpretation of interaction experiments.

This work was supported by the National Science Foundation through
Grant Number DMR-9730189 and by the Deutsche Forschungsgemeinschaft.


\end{document}